\documentclass[12pt]{article}

\newcounter{multieqs}

\newcommand{\bq}{\begin{equation}}
\newcommand{\fq}{\end{equation}}
\newcommand{\bqr}{\begin{eqnarray}}
\newcommand{\fqr}{\end{eqnarray}}

\newcommand{\be}{\begin{equation}}
\newcommand{\ee}{\end{equation}}
\newcommand{\eq}[1]{(\ref{#1})}

\newcommand{\bm}[1]{\mbox{\boldmath $#1$}}

\newcommand{\hoch}[1]{$\, ^{#1}$}

\def\bd{\begin{document}}
\def\ed{\end{document}}
\def\nn{\nonumber}
\def\bea{\begin{eqnarray}}
\def\eea{\end{eqnarray}}
\let\bm=\bibitem
\let\la=\label

\def\npb#1#2#3{Nucl. Phys. {\bf{B#1}} #3 (#2)}
\def\plb#1#2#3{Phys. Lett. {\bf{#1B}} #3 (#2)}
\def\prl#1#2#3{Phys. Rev. Lett. {\bf{#1}} #3 (#2)}
\def\prd#1#2#3{Phys. Rev. {D \bf{#1}} #3 (#2)}
\def\cmp#1#2#3{Comm. Math. Phys. {\bf{#1}} #3 (#2)}
\def\cqg#1#2#3{Class. Quantum Grav. {\bf{#1}} #3 (#2)}
\def\nppsa#1#2#3{Nucl. Phys. B (Proc. Suppl.) {\bf{#1A}}#3 (#2)}
\def\ap#1#2#3{Ann. of Phys. {\bf{#1}} #3 (#2)}
\def\ijmp#1#2#3{Int. J. Mod. Phys. {\bf{A#1}} #3 (#2)}
\def\rmp#1#2#3{Rev. Mod. Phys. {\bf{#1}} #3 (#2)}
\def\mpla#1#2#3{Mod. Phys. Lett. {\bf A#1} #3 (#2)}
\def\jhep#1#2#3{J. High Energy Phys. {\bf #1} #3 (#2)}
\def\atmp#1#2#3{Adv. Theor. Math. Phys. {\bf #1} #3 (#2)}

\newcommand{\EQ}[1]{\begin{equation} #1 \end{equation}}
\newcommand{\AL}[1]{\begin{subequations}\begin{align} #1 \end{align}\end{subequations}}
\newcommand{\SP}[1]{\begin{equation}\begin{split} #1 \end{split}\end{equation}}
\newcommand{\ALAT}[2]{\begin{subequations}\begin{alignat}{#1} #2 \end{alignat}\end{subequations}}
\def\beqa{\begin{eqnarray}} 
\def\eeqa{\end{eqnarray}} 
\def\beq{\begin{equation}} 
\def\eeq{\end{equation}} 

\def\N{{\cal N}}
\def\sst{\scriptscriptstyle}
\def\thetabar{\bar\theta}
\def\Tr{{\rm Tr}}
\def\one{\mbox{1 \kern-.59em {\rm l}}}

\def\a{\alpha}		\def\da{{\dot\alpha}}
\def\b{\beta}		\def\db{{\dot\beta}}
\def\c{\gamma} 	\def\C{\Gamma}	\def\cdt{\dot\gamma}
\def\d{\delta}	\def\D{\Delta}	\def\ddt{\dot\delta}
\def\e{\epsilon}		\def\vare{\varepsilon}
\def\f{\phi}	\def\F{\Phi}	\def\vvf{\f}
\def\h{\eta}
\def\k{\kappa}
\def\l{\lambda}	\def\L{\Lambda}
\def\m{\mu}	\def\n{\nu}
\def\p{\pi}	\def\P{\Pi}
\def\r{\rho}
\def\s{\sigma}	\def\S{\Sigma}
\def\t{\tau}
\def\th{\theta}	\def\Th{\Theta}	\def\vth{\vartheta}
\def\X{\Xeta}
\def\z{\zeta}

\def\cA{{\cal A}} \def\cB{{\cal B}} \def\cC{{\cal C}}
\def\cD{{\cal D}} \def\cE{{\cal E}} \def\cF{{\cal F}}
\def\cG{{\cal G}} \def\cH{{\cal H}} \def\cI{{\cal I}}
\def\cJ{{\cal J}} \def\cK{{\cal K}} \def\cL{{\cal L}}
\def\cM{{\cal M}} \def\cN{{\cal N}} \def\cO{{\cal O}}
\def\cP{{\cal P}} \def\cQ{{\cal Q}} \def\cR{{\cal R}}
\def\cS{{\cal S}} \def\cT{{\cal T}} \def\cU{{\cal U}}
\def\cV{{\cal V}} \def\cW{{\cal W}} \def\cX{{\cal X}}
\def\cY{{\cal Y}} \def\cZ{{\cal Z}}

\def\ua{\underline{\alpha}}
\def\ub{\underline{\phantom{\alpha}}\!\!\!\beta}
\def\uc{\underline{\phantom{\alpha}}\!\!\!\gamma}
\def\um{\underline{\mu}}
\def\ud{\underline\delta}
\def\ue{\underline\epsilon}
\def\una{\underline a}\def\unA{\underline A}
\def\unb{\underline b}\def\unB{\underline B}
\def\unc{\underline c}\def\unC{\underline C}
\def\und{\underline d}\def\unD{\underline D}
\def\une{\underline e}\def\unE{\underline E}
\def\unf{\underline{\phantom{e}}\!\!\!\! f}\def\unF{\underline F}
\def\unm{\underline m}\def\unM{\underline M}
\def\unn{\underline n}\def\unN{\underline N}
\def\unp{\underline{\phantom{a}}\!\!\! p}\def\unP{\underline P}
\def\unq{\underline{\phantom{a}}\!\!\! q}
\def\unQ{\underline{\phantom{A}}\!\!\!\! Q}
\def\unH{\underline{H}}

\def\As {{A \hspace{-6.4pt} \slash}\;}
\def\Ds {{D \hspace{-6.4pt} \slash}\;}
\def\ds {{\del \hspace{-6.4pt} \slash}\;}
\def\ss {{\s \hspace{-6.4pt} \slash}\;}
\def\ks {{ k \hspace{-6.4pt} \slash}\;}
\def\ps {{p \hspace{-6.4pt} \slash}\;}
\def\pas {{{p_1} \hspace{-6.4pt} \slash}\;}
\def\pbs {{{p_2} \hspace{-6.4pt} \slash}\;}

\def\Fh{\hat{F}}
\def\Xh{\hat{X}}
\def\ah{\hat{a}}
\def\xh{\hat{x}}
\def\yh{\hat{y}}
\def\ph{\hat{p}}
\def\xih{\hat{\xi}}

\def\psit{\tilde{\psi}}
\def\Psit{\tilde{\Psi}}
\def\tht{\tilde{\th}}
 
\def\At{\tilde{A}}
\def\Qt{\tilde{Q}}
\def\Rt{\tilde{R}}

\def\ft{\tilde{f}}
\def\pt{\tilde{p}}
\def\qt{\tilde{q}}
\def\vt{\tilde{v}}

\def\delb{\bar{\partial}}
\def\bz{\bar{z}}
\def\Db{\bar{D}}

\def\d{\delta}\def\D{\Delta}\def\ddt{\dot\delta}

\def\pa{\partial} \def\del{\partial}
\def\xx{\times}

\def\trp{^{\top}}
\def\inv{^{-1}}
\def\dag{{^{\dagger}}}\def\pr{^{\prime}}

\def\rar{\rightarrow}
\def\lar{\leftarrow}
\def\lrar{\leftrightarrow}

\newcommand{\0}{\,\!}      
\def\one{1\!\!1\,\,}
\def\im{\imath}
\def\jm{\jmath}

\newcommand{\tr}{\mbox{tr}}
\newcommand{\slsh}[1]{/ \!\!\!\! #1}

\def\vac{|0\rangle}
\def\lvac{\langle 0|}

\def\hlf{\frac{1}{2}}
\def\ove#1{\frac{1}{#1}}

\def\Box{\square}
\def\ZZ{\mathbb{Z}}
\def\CC#1{({\bf #1})}
\def\bcomment#1{}
\def\bfhat#1{{\bf \hat{#1}}}
\def\VEV#1{\left\langle #1\right\rangle}

\newcommand{\ex}[1]{{\rm e}^{#1}} \def\ii{{\rm i}}

\begin{document}
 

\hfill{HUB-EP-01/40}

\hfill{hep-th/0110147}

\vspace{20pt}
 
\begin{center}
 
{\Large \bf Noncommutative $SU(N)$ and Gauge Invariant Baryon Operator }
\vspace{30pt}
 
{\large Chong-Sun Chu\hoch{1} and Harald Dorn\hoch{2}}
 
\vspace{15pt}
\begin{itemize}
\item[$^1$]
{\small \em Centre for Particle Theory, Department of Mathematical Sciences,
University of Durham, \\ Durham, DH1 3LE, UK}
 
\item[$^2$]
{\small \em Institut f\"{u}r Physik, Humboldt Universit\"{a}t zu Berlin,
Invalidenstrasse 110, \\ D-10115   Berlin, Germany}
 
\end{itemize}
{\bf Abstract}
 
\end{center}
\noindent
We propose a constraint on the noncommutative gauge theory with $U(N)$ gauge group 
which gives rise to a noncommutative version of the $SU(N)$  gauge
group. The baryon operator is also constructed.
   

\newpage 

\section{Introduction}

Noncommutative geometry of the form
\be \label{ncr}
[x^\m,x^\n] = i \th^{\m\n},
\ee
has got a lot of interest recently. (See \cite{sw,dn,szabo} for
a comprehensive introduction and an extensive list of references.) 
Part of the reason is because it
appears in a certain corner of moduli space of string \cite{dh,CH1,vs}
and M theory \cite{cds,CHL} and so cannot be ignored. 

Noncommutative gauge theory with gauge group $U(N)$ has been
constructed and analysed quite extensively in the literature. It was
first pointed out in \cite{sw} that there is an obstacle in 
the naive way to construct 
noncommutative gauge theory with gauge group other than $U(N)$. Since
then there had been a number of proposals \cite{sj1,wess1,sj2,wess2} 
to construct  noncommutative gauge theory with gauge group different from 
$U(N)$.

In this letter we propose a construction of noncommutative $SU(N)$
gauge theory.  The construction follows 
similar ideas as in \cite{sj1,sj2} by imposing a constraint on the
noncommutative $U(N)$ gauge configurations. The constraint selects out the
corresponding gauge configurations that we propose to be identified as
noncommutative $SU(N)$ configurations.
We also construct a gauge invariant baryon operator.

\section{Noncommutative $SU(N)$}

Consider a noncommutative gauge theory with gauge group $U(N)$. The
action is
\be
S = \frac{1}{4} \int d x \; {\rm Tr} F_{\m \n} *  F_{\m \n}~,
\ee
where
\be
F_{\m\n} = \del_\m A_\n - \del_\n A_\m + i g (A_\m * A_\n -A_\n *
A_\m)    
\ee
and $g $ is the gauge coupling. 
The gauge transformation is given by
\be \label{At}
A_{\mu}(x) \rightarrow U(x) * A_{\mu}(x) * U(x)^{\dagger}
-\frac{i}{g} U(x) * \partial_{\mu} U(x)^{\dagger}, 
\ee
where $U(x) \in U(N) $  with $U(x) * U(x)^{\dagger}=U(x)^\dag * U(x) =1$.

The Wilson line in  noncommutative gauge theories is defined by
\be W(x,C)= P_* \exp \left(i g \int_0^1 d \sigma {\frac{d \z ^{\mu}}{d \sigma}} A_{\mu}(x+\zeta(\sigma))\right)  , \ee
where $C$ is the curve
\be \label{cont}
C = \{ \z^{\mu}(\sigma), 0\leq \s \leq 1 \; | \; \zeta(0) = 0,  \zeta(1)=l \},
\ee
and  $P_{*}$ is the
path ordering with respect to the star product
\be W(x,C)= \sum_{n=0}^{\infty} (ig)^n
\int_0^1 d\sigma_1 \int_{\sigma_1}^1 d \sigma_2 ...
\int_{\sigma_{n-1}}^1 \!\!\!\!\!\! d \sigma_n \
\z^{'}_{\mu_1}(\s_1) ... \z^{'}_{\mu_n}(\s_n)
A_{\mu_1}(x+\z(\s_1))* ... * A_{\mu_{n}}(x+\z(\s_n)).  \ee
It is easy to verify that $W(x,C)$ transforms under a gauge transformation as
\be
W(x,C) \to U(x) * W(x,C) * U^\dag (x+l) ~ .
\ee
The open Wilson line is an important building block for constructing
gauge invariant operators \cite{rey,gross}. The crucial observation is that in
noncommutative geometry \eq{ncr}, the plane wave $e^{i k x}$ is a
translational operator
\be
e^{i k x} * f(x) = f(x+ \th k) * e^{i k x},
\ee 
therefore one can construct gauge invariant operators with the help of
the  open Wilson line. Let $\cO(x)$ be an operator transforming in the
adjoint (e.g. ${\rm Tr} F^n$), i.e.
\be
\cO (x) \rightarrow  U(x) *\cO (x)*U^{\dag}(x) ~,
\ee
then one can introduce 
\be \label{ncfour}
\tilde{\cO} (k) := \int d x \; \cO(x) * W(x,C_k) * e^{i k x},
\ee
where the subscript $k$ of $C_k$ denotes the possible $k$
dependence of the contour. 
$\tilde{\cO} (k)$ is a generalization of the Fourier transform
of the operator $\cO$ to the noncommutative case.  
It reduces to the usual Fourier transform in the commutative limit. The tilde 
reminds us that $\tilde{\cO} (k)$ is
not exactly the usual Fourier transform of $\cO(x) $.
It is easy to show that the following momentum space operator
\be
{\rm Tr} \;\tilde{\cO} (k),
\ee
is gauge invariant
if  $C_k$ satisfies the condition
\be \label{lk}
l^\m = \th^{\m \n}k_\n .
\ee
Although (\ref{lk}) is sufficient to guarantee gauge invariance, straight
contours play a special role. Then the insertion point for $\cO$ on $C_k$
is arbitrary \cite{gross} and in addition one has the remarkable
identity \cite{DW,leigh}
\be
 e_*^{ik(x-g\theta A(x))}~=~W(x,C_k)*e^{ikx}~.
\ee
The subscript $*$ on the l.h.s. indicates that the exponential is understood
as a power series of the star multiplied exponent. 
The combination $x-g\theta A(x) $ is just the covariant coordinate in the sense
of \cite{peter}.
From now on we always choose straight contours.

To construct noncommutative gauge theory with gauge group $SU(N)$, we 
can try to follow the approach of \cite{sj1,sj2} by imposing 
constraints on the gauge configurations $A$ and gauge
transformation parameters of the noncommutative $U(N)$.
For finding a suitable constraint to fix a noncommutative version of 
$SU(N)$ (we denote it from now by $ncSU(N)$ ) it is helpful to recall the 
reason why simple tensoring
$SU(N)$ with the star product fails. If one imposes ${\rm Tr}\lambda (x)=0$
to single out the modified $ncSU(N)$ Lie algebra,
it turns out that this condition due to the non-commutativity of the star
product does not close, i.e. ${\rm Tr}[\lambda (x) ,\mu (x)]_{*}\neq 0$.
A formulation of the same problem in terms of $U(N)$ group elements
would impose the vanishing trace condition on the Maurer-Cartan form, i.e.
${\rm Tr}(U^{\dagger}*dU)=0$. Since ${\rm Tr}((U*V)^{\dagger}*d(U*V))=
{\rm Tr}(V^{\dagger}*U^{\dagger}*dU*V)+{\rm Tr}(U^{\dagger}*dU)$, we see
again that the lack of cyclic invariance under the matrix trace prevents
$U*V$ to fulfil the constraint if $U$ and $V$ do separately. Cyclic
invariance is restored if the total trace with respect to internal
indices and the spacetime points is taken. Now certainly 
$\int dx {\rm Tr}(U^{\dagger}*dU)=0$ is too weak and one should look for a 
localized version, i.e. in the language
of Fourier transforms for an extension from momentum $k=0$ to generic
$k\neq 0$. This problem is similar to the one described above in connection
with the construction of gauge invariant quantities. 

Motivated by these remarks we impose the following constraint on the allowed
gauge field configurations:
\be \label{c1}
{\rm Tr}\; \At(k)~ := ~\int dx \; {\rm Tr}(A(x) * W(x,C_k)) * e^{i k x}~=~ 0~,
~~~~~\forall k ~.
\ee
The constraint \eq{c1} is a condition on the allowed gauge
configurations $A$ of $U(N)$ that  can be identified as $ncSU(N)$
configurations. It is the generalization of the traceless condition for the 
commutative $SU(N)$ gauge
fields. Under a gauge transformation, it transforms as
\be
{\rm Tr}\; \At(k)\rightarrow 
{\rm Tr}\; \At(k)~+~\frac{i}{g}\int dx {\rm Tr}(U^{\dagger}(x)
*dU(x)*W(x,C_k))*e^{ikx} ~.
\ee
So in order for \eq{c1} to be gauge invariant, we need to impose the
condition 
\be \label{c2}
\int d x \; {\rm Tr}\;( U^{\dagger}(x) * dU(x)* W(x,C_k)) * e^{ikx} =0~,~~~~~
\forall k~,
\quad \mbox{for  $x$-dependent  $U$,}
\ee
on the allowed gauge transformations. Note that the allowed gauge
transformation $U(x)$ is generally a  gauge field dependent gauge
transformation. Strictly speaking we
should write $U^A$. In the following we will drop the superscript and simply
write $U$. This is to be
distinguished from the case of noncommutative $U(N)$.
In the commutative limit, \eq{c2} reduces to the usual traceless 
condition ${\rm Tr} (d\l ) =0$ where $U = \exp {i\l} \in SU(N)$.
For $x$-independent
gauge transformations, the condition \eq{c2} gives no extra
information and it is natural to consider gauge transformations that
are traceless
\be \label{c3}
{\rm Tr}\; \l~=~0, \quad \mbox{for  $x$-independent  $U=e^{i\l}$}.
\ee
Thus we propose that 
\eq{c1},\eq{c2} and \eq{c3} together  provide a 
characterization of noncommutative $ncSU(N)$ gauge configurations.

Furthermore, we note the following composition law for our $ncSU(N)$. 
Denote the constraint (\ref{c2}) as $f(U,A)=0$ and as $ A^U$
the gauge transform of $A$ according to (\ref{At}), then one has the composition
law
\be
f(V*U,A)~=~f(V,A^U)~+~f(U,A)~
\ee
for $ncSU(N)$.
This ensures the consistency of imposing (\ref{c1}) and (\ref{c2}) under
successive gauge transformations.

Before closing this section we want to comment on the issue of
nontrivial solutions for our constraints. Both constraints are understood
to be imposed for any $k$. Therefore, an equivalent form of (\ref{c1}) and 
(\ref{c2}) which no longer contains $k$ would be highly welcome. 
For the commutative case $\theta =0$ the Wilson line $W$ is absent and we 
have the situation of standard Fourier transformation. Vanishing of the
Fourier transform of ${\rm Tr}A$ for all $k$ is equivalent to ${\rm Tr}A=0$
for all $x$. Our aim is to get for $\theta\neq 0$ a similar pure coordinate
space constraint. To this goal let us expand $W$ and further Taylor expand
the appearing $A(x+\sigma\theta k)$. All arising factors of $k$ can be
thought as generated by differentiations of $e^{ikx}$. Then these 
differentiations, by partial integrations, will be moved to the remaining 
factors in the $x$-integral. Under this integral the $*$ in front of
$e^{ikx}$ can be dropped and we arrive at the standard Fourier transformation
of an infinite power series in $A,~\theta ,~\partial $. Performing
these manipulations explicitly we find up to $O(\theta ^3)$
\bea
{\rm Tr}\left ( A_{\nu}-g(\theta\partial )^{\mu}(A_{\nu}*A_{\mu})-
\frac{i}{2}g(\theta\partial )^{\mu _1}(\theta\partial )^{\mu _2}
(A_{\nu}*\partial _{\mu _1}A_{\mu _2})\right .~~~~~~~~~~~~~~~~~~~~~~~~~~~~~
\nonumber\\
~~~~~~~~~~~~~~~~~~~~~~\left .+\frac{1}{2}g^2 (\theta\partial )^{\mu _1}
(\theta\partial )^{\mu _2}(A_{\nu}*A_{\mu _1}*A_{\mu _2})\right )~+~
O(\theta ^3)~=~0~.
\eea
The corresponding equivalent to (\ref{c2}) is obtained by
replacing $A_{\nu}$ by $U^{\dagger}\partial _{\nu}U$ and keeping the $A_{\mu}$.
While obviously all finite order approximations have nontrivial solutions,
at this level of discussions it is far from obvious whether the infinite
power series allows for solutions.

However one can nevertheless find an argument for the existence of
nontrivial solutions. Let us define (A similar construction
for quantities transforming in the adjoint has been used in \cite{leigh}.)
\be
\hat A(y)~=~\frac{1}{(2\pi )^D}\int dk~e^{-iky}~\tilde A (k)
\ee
and similar for $U^{\dagger}\partial _{\nu}U$. Then ${\rm Tr}\tilde A(k)=0$
for all $k$ is equivalent to ${\rm Tr}\hat A(y)=0$ for all $y$. The
map $A\rightarrow\hat A$ is a map of coordinate space functions. For
$\theta =0$ it is the identity map hence invertible. Assuming that continuity
ensures invertibility also for $\theta\neq 0$ we have nontrivial solutions
of our constraints for free.

The vacuum configuration $A=0$ has a distinguished position. Then $W=1$ and
 \eq{c2}
says the admissible $ncSU(N)$  gauge transformations satisfy
\be
\int dx \; {\rm Tr}\; ( U^{\dagger}(x) * dU(x))* e^{ikx} =0~, \quad \forall k~.
\ee
This is equivalent to ${\rm Tr}\; ( U^{\dagger}(x) * dU(x))=0$ and
 on substituting to \eq{At} implies that then $A= i dU * U^{\dagger} /g$,
is also a $ncSU(N)$ configuration. 

\section{Gauge invariant baryon operator}

We start with a commutative gauge theory with a color gauge group $SU(N)$ 
\footnote{We suppress flavor and spin indices.}.
Let $q^i$ be a set of fermionic fields in the  
fundamental representations of $SU(N)$. 
They transform under $SU(N)$ as
\be
q^{i} \rightarrow  U^i_jq^{j}, \quad U \in SU(N)~. 
\ee
One can introduce the operator
\be
M^i{}_j = q^{i} \bar{q}_{j}.
\ee
It transforms as
\be
M \rightarrow  U~M~U^{\dag}~.
\ee
A set of $SU(N)$ gauge invariant operators can be constructed from
powers of $M$ as 
\be
{\rm Tr}\; {M}^n, \quad n = 1,2,\cdots.
\ee
In addition, the determinant ${\rm Det}M$ can be related to traces of 
powers of $M$ using the formula
\be \label{det}
{\rm Det} M = \sum _{n_1+2n_2+...+Nn_N=N}c^{(N)}_{n_1n_2...n_N}
({\rm Tr}M)^{n_1}({\rm Tr}M^2)^{n_2}...({\rm Tr}M^N)^{n_N}~,
\ee
which e.g. for $N=2,3$ means
\bea
{\rm Det} M&=&-\frac{1}{2}\left ({\rm Tr}M^2~-~({\rm Tr}M)^2\right )~,
~~~N=2\nonumber\\
{\rm Det} M&=&\frac{1}{3}{\rm Tr}M^3~-~\frac{1}{2}{\rm Tr}M{\rm Tr}M^2~+~
\frac{1}{6}({\rm Tr}M)^3~,~~~N=3~.
\eea
In the commutative case, the standard gauge invariant baryon operator is 
given by
\be \label{Bdef}
B(x) = \frac{1}{N!} \epsilon_{i_1 i_2 \cdots i_N} q^{i_1}(x) \cdots q^{i_N}(x)~.
\ee
The magnitude of this 
baryon operator $B$ is related to ${\rm Det} M$ through the formula
\be \label{BBM}
{\rm Det} M~=~N!~B B^\dagger ~.
\ee
Up to now we have related the absolute value of the baryon operator
to traces of powers of $M$, which transform in the adjoint.

This can be used as a starting point for the definition of the square
of the absolute value of a baryon operator in the noncommutative case.
We know how to form gauge invariant quantities out of operators transforming 
in the adjoint. Therefore we define
\be
N!B B^\dagger (y) :=\sum _{n_1+2n_2+...+Nn_N=N}c^{(N)}_{n_1n_2...n_N}
({\rm Tr}\hat M(y))^{n_1}({\rm Tr}\widehat{M^2}(y))^{n_2}...({\rm Tr}\widehat 
{M^N}(y))^{n_N}~,
\ee
with
\be
\widehat{M^j}(y)~=~\frac{1}{(2\pi )^D}\int dk~e^{-iky}~\widetilde{M^j}(k)~.
\ee
The quantity defined by (26) is gauge invariant under noncommutative $U(N)$
and reproduces $N!B B^\dagger (y)$ with $ B(y)$ given by (\ref{Bdef}) in
the commutative limit. The phase of $B$ remains undetermined in the 
construction just presented. No use has been made of the $ncSU(N)$ constraint.

To proceed with a different construction, 
we consider Wilson lines for contours $C$ running to infinity, in particular
$W(x,C_{\infty})$ with
\be
C_{\infty} = \{ \z^{\mu}(\sigma), 0\leq \s \leq 1 \; | \; \zeta(0) = \infty,  
\zeta(1)=0 \}~.
\ee
They transform under the gauge transformation \eq{At} as
\be
W(x,C_{\infty})\rightarrow U(\infty )*W(x,C_{\infty})*U^{\dagger}(x)~ .
\ee
For gauge transformation that are trivial at infinity, i.e.  $U(\infty) =1$,
this becomes
\be \label{W-anti}
W(x,C_{\infty})\rightarrow  W(x,C_{\infty})*U^{\dagger}(x) ,
\ee
which effectively is a transformation in the anti-fundamental representation.
Using $W(x,C_{\infty})$, one can construct the manifestly gauge invariant
combination $W*q$ and use it as the building block for a gauge
invariant baryon operator. We define in the $x$-space the following operator
\be \label{cB}
\cB (x) = \frac{1}{N!} \e_{i_1 \cdots i_N} \; (W^{i_1}{}_{j_1} * q^{j_1})
*\cdots *(W^{i_N}{}_{j_N} * q^{j_N}) ~.
\ee
Note that $\cB$ is manifestly gauge invariant for all noncommutative $U(N)$
transformations approaching the identity at infinity, and not just for
$ncSU(N)$ ones.
However, one should nevertheless restrict oneself to noncommutative
$SU(N)$ configuration in the definition \eq{cB} of $\cB$. 
Indeed in the commutative limit, 
\be
\cB(x) =  {\rm Det} ~W(x,C_{\infty}) \cdot B(x) 
\ee
where $B$ given by \eq{Bdef} is the usual baryon operator. 
Thus  $\cB =  B$ only if $A$ is in commutative 
$SU(N)$,  since  then 
\be \label{detW}
{\rm Det} ~W(x,C_{\infty})=1, \quad \mbox{in the commutative limit}.
\ee  
Just this limiting property is realised if the $ncSU(N)$ constraint \eq{c1}
is imposed.

We also note that our baryon $\cB$ is invariant 
with respect to the rigid, i.e. $x$-independent, $SU(N)$ gauge
transformations since constants can be factored out of the star
products. 

Finally we remark that the construction in this section by parallel
transporting the quarks from infinity works so long as the Wilson loop
in anti-fundamental representation \eq{W-anti} can be constructed. The
correct commutative limit is guaranteed by \eq{detW}. For example, our
construction here can be applied equally well to \cite{wess1,wess2}.

\section{Discussions} 

In this paper, we have proposed a definition of noncommutative
$SU(N)$. 
We would like to comment on the relation of our work to other approaches
to the construction of noncommutative gauge theories beyond $U(N)$.
Working with enveloping algebra valued gauge fields whose components
are functions of standard Lie algebra valued fields, noncommutative
gauge theories have been constructed for arbitrary Lie algebras 
\cite{wess1,wess2}. The construction is based on the use of the 
Seiberg-Witten map.
For practical calculations this map is treatable as a power expansion in 
$\theta $ only. Since our main motivation was the construction of
physical relevant gauge invariant quantities in all orders of $\theta$,
we followed another approach, namely the use of suitable constraints
within noncommutative $U(N)$. Our constraint for
$ncSU(N)$ differs qualitatively from those used for the noncommutative
versions of $O(N)$ and $Usp(2N)$ \cite{sj1,sj2}. The new feature
is the dependence of the gauge transformation constraint on the gauge
field. Thinking in terms of covariant coordinates \cite{peter}, this 
seems to be a quite  generic feature in noncommutative
geometry.

Just this property reminds one a little bit on the dependence of the 
noncommutative gauge transformation both on the commutative one and
the commutative gauge field within the Seiberg-Witten map. Nevertheless
a sketchy look at the $U(N)$ SW map indicates that, to lowest order
in $\theta$, the images of commutative $SU(N)$ fields do not
necessarily obey our constraint. This would imply that our noncommutative
version of $SU(N)$ is different from the one constructed along the lines of
\cite{wess1,wess2}.

The consideration in this paper is mostly a classical one. 
Our constraint is motivated purely from the field theory side.
It is possible that the noncommutative $SU(N)$ gauge theory is an
effective description of a certain string construction. It would also
be interesting  to see if there is a dual description in terms
of gravity. An interesting related issue is to
understand the existence of the baryon vertex operator from the
AdS/CFT point of view \cite{gross,go,witten}. 
However these are
completely open for the moment. 

On the other hand, one may try to
think of  the noncommutative $SU(N)$ gauge theory as a quantum theory
and try to analyse its quantum properties. As a first step, one needs a 
correct implementation of the constraint at the
quantum level, for example using the Dirac quantization.
It would be interesting to perform an one loop analysis similar to
those in \cite{armoni,bonora}
and clarify its relation to that for noncommutative $U(N)$.

On the more phenomenological level, it may be interesting to adopt 
our construction of the noncommutative $SU(N)$ and 
the baryon operator in studying
standard model with noncommutative $SU(3) \times SU(2) \times U(1)$ 
gauge symmetry.

It would also be
interesting to generalize our construction to include matters
transforming in other nontrivial representations other than the
fundamental representation of the noncommutative
$SU(N)$ by imposing an appropriate set of constraints. 

\section*{Acknowledgments} 
C.S.C. would like to thank the warm hospitality of 
the theoretical physics group of the Humboldt University 
where this work was initiated. We would like to thank Douglas Smith
for discussions.

\end{document}